\documentclass[letterpaper,10pt]{article}
\usepackage{osameet2}

\usepackage{amsmath,amssymb}

\usepackage[dvips,colorlinks=true,bookmarks=false,citecolor=blue,urlcolor=blue]{hyperref} 
\usepackage{graphicx}
\usepackage{bm}
\usepackage{color}
\def\be{\begin{equation}}
\def\ee{\end{equation}}
\def\bea{\begin{eqnarray}}
\def\eea{\end{eqnarray}}

\def\ra{\rangle}

\begin{document}

\title{Uncollapsing the wavefunction}
\author{Andrew N. Jordan}
\address{Department of Physics and Astronomy, University of Rochester, Rochester, New York 14627, USA}
\email{jordan@pas.rochester.edu}
\author{Alexander N. Korotkov}
\address{ Department of Electrical Engineering, University of California,
Riverside, CA 92521-0204, USA}
\email{korotkov@ee.ucr.edu}

\begin{abstract}
The undoing of quantum measurements is discussed in the broader context of irreversibility in physics.  We give explicit examples of how a wavefunction can be uncollapsed in two solid-state experimental set-ups.   Wavefunction uncollapse shows the quantum observer paradox in a new and significantly clearer way.
\end{abstract}


\section{Introduction}

\begin{figure}[b]
  \centering
 \includegraphics[width=14cm]{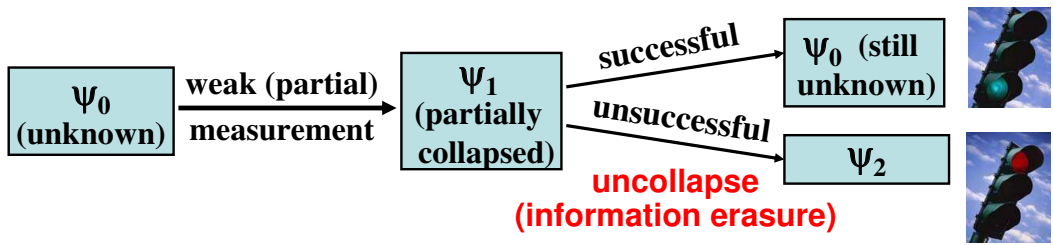} 
\caption{Diagram indicating the structure of the quantum undemolition (QUD) measurement.  Starting with a completely unknown quantum state, a weak measurement is made, partially collapsing the state of the qubit.  A second weak measurement can be made (the details of which must be tailored to the system in question) that can uncollapse the wavefunction, fully restoring the original (unknown) state if the information from the first measurement is erased.  This information erasure is stochastic, so it may be successful or unsuccessful in a given attempt.  The physical detector will unambiguously indicate success (a green light), or failure (a red light). } 
\label{fig1}
\end{figure}
The presence of irreversibility in nature has historically been a perplexing one.   Starting from the observation that all known fundamental laws of physics are reversible in time (Newton's laws, Schr\"odinger's equation, Einstein's equations, etc.), the questions is natural:  whence irreversibility?  Broadly speaking, there seems to be two schools of thought.  The first school is that irreversibility is only effective.  It originates from a lack of information, an inevitable coarse-graining of the physics.  Such an approach is embodied by Loschmidt's comment to Boltzmann, that given a collection of atoms that are localized in one part of container which are then released and allowed to spread throughout the container (thus increasing the entropy) for a time $t$, if the velocity of all the atoms are then reversed, they will come back to the initial ordered state after another time $t$, to which Boltzmann replied, ``it is {\it you} who would invert the velocities'' \cite{boltz}.
More modern approaches to irreversibility (even in strictly classical systems) have resolved the issue through the question of allowable observables.  By restricting observables to a set of functions that are mathematically smooth, it can be rigorously demonstrated for chaotic systems that any correlation function of these observables decays both forward and backward in time \cite{ruelle}.

In the quantum realm, analogous schools of thought are present.  For example, the ``Loschmidt  echo'' (closely related to fidelity) has been introduced as a figure of merit for the degree of quantum chaos of a system \cite{peres}.  The wavefunction is evolved in time under one Hamiltonian, and then time-reversed with a slightly different Hamiltonian.  Generally speaking, the longer the initial time evolution, the smaller the overlap with the original state after the time reversal, showing irreversibility to be the product of small imprecisions that are amplified by the chaotic dynamics. 

For the other school of thought, irreversibility is fundamental.   Recalling the reversibility of the Schr\"odinger equation, the only place to introduce fundamental irreversibility is in the phenomena of quantum measurements.  It is well known that the process of measurement cannot be described by unitary operations, and are irreversible.  The foundations of quantum measurement theory are still controversial, despite the intervening years since quantum theory was discovered.  Here, we shall briefly describe two approaches to quantum measurement, and how irreversibility arises. 

We first discuss the text-book treatment of measurement in terms of projections of a state onto the eigenstate of an observable.  The standard expression is the Born rule $(P_k = |\psi_k|^2)$, describing the probability of observing a particular result \cite{Neumann}.  When this result is registered, the quantum state is altered to be the eigenstate of that observed eigenvalue.  This rule is an axiom of the standard Copenhagen interpretation, and any kind of explanation of the origins of the rule are typically ignored.  Later commentary describes an interesting link between the collapse of the wavefunction and the making of a measurement record.  For example, the last page of my undergraduate textbook states  ``However, the making of a record is essentially an irreversible process; the record is indelible.  No process exists that will undo a measurement and delete a record, substituting another.'' \cite{book} .   Another example is John Wheeler's famous article, {\it Law without law}, in which he states  ``We are dealing with [a quantum] event that makes itself known by an irreversible act of amplification, by an indelible record, an act of registration.'' \cite{wheeler} . Somehow, the leaving of tell-tale tracks in the physical space and time introduces irreversibility into the quantum world at a fundamental level.  

Another common approach to resolving the measurement problem is the decoherence theory of quantum measurement, popularized by Zurek and collaborators \cite{zurek}.  In this theory, the presence of the measuring device acts as a bath, continuously dephasing the originally coherent state to a fully mixed density matrix.  Here, irreversibility enters because the information about the complete quantum state is jumbled up inside the quantum environment which cannot be accessed.  Any attempt to rephase the system is doomed to fail for the same reasons that the Loschmidt-echo thought experiment cannot recover the overlap in a quantum chaotic system:  it is not feasible to microscopically reverse the dynamics of the macroscopic dynamics, just as it is not feasible to get a gas to localize into the corner of the box by reversing all of the velocities.    If one measurement by a macroscopic measuring device dephases the system a little, a second measurement will only serve to further dephase the system.  Therefore, saving (an unfeasible) microscopic time reversal of all elementary particles participating in the problem, a dephasing measurement process is irreversible.

Lurking underneath this discussion is a conflict between two fundamentally different approaches to quantum measurement.  In one class of approaches, the measurement process is fundamentally a mechanistic one, where wavepacket reduction is governed by physical laws.  The decoherence program is one such attempt where the mechanism is ultimately reducible to random forces of the measuring device on the quantum system.  The so-called ``GRW'' model is another such attempt of reformulating wavefunction collapse of a single quantum measurement in terms of stochastic forces outside of standard quantum theory \cite{GRW}.  An alternative approach is to view the quantum state in a Bayesian framework, where the quantum state reflects our knowledge of the system.  From a philosophical point of view, this analysis of continuous quantum measurements is quite similar to the Copenhagen interpretation of projective measurements.
In such a framework, wavefunction reduction is simply an updating of our knowledge based on new information we receive.  Our knowledge can be updated abruptly - as in the case of wavefunction collapse, or it can be updated gradually - as in the case of continuous measurement.   

The purpose of the present paper is to advance the consequences of the Bayesian approach to quantum measurement, specifically regarding the (ir)reversibility issue.   Our central claim is that quantum measurements can be undone \cite{us}, but only in the case where there is incomplete information about the state, a so-called {\it weak} measurement \cite{weak-meas}.  While this claim contradicts the above two more conventional approaches to quantum measurement, the seed of the idea can be seen as a loophole in the above quotations.  Indeed, if there were some way to erase the informative part of the measurement record, this would nullify the information obtained in the first measurement, thereby undoing its effect on the quantum system.   

The sequence of partial collapse and uncollapse is shown in Fig. 1.  Starting with an unknown initial state $\psi_0$, the first weak measurement registers a result, partially collapsing the wavefunction into another unknown state $\psi_1$.  To undo this operation (mathematically described as a one-to-one map), a second undoing measurement is applied that is formally the inverse of the first weak measurement operation.  Application of the inverse operation is a stochastic process, so uncollapsing the wavefunction will sometimes succeed (in which case, the original unknown wavefunction $\psi_0$ is fully restored), and sometimes fail (producing some other state $\psi_2$).  When the undoing procedure succeeds, the (classical) information obtained from the first measurement must be erased by the second measurement.
We refer to this as a Quantum Undemolition (QUD) measurement.  Notice the QUD procedure has nothing necessarily to do with time-reversal.  Indeed, the explicit examples we give will write and erase the information with {\it different} physical processes.
 
Before proceeding to specific systems, we note that the issue of undoing a quantum measurement has been raised previously (see \cite{Ueda-99,Ban-01,D'Ariano-03} and related papers \cite{reversib}).  In particular, we briefly discuss the relation of the theory here to the ``quantum eraser'' of Scully and Dr\"uhl \cite{Scully-82}. In this proposal, the which-path information of a `two-slit' photon was stored in the quantum state of an atom. If this and similar photons were allowed to form a pattern on a detection screen, there would be no interference.  However, if this which-path information were erased (even after the photon passed through the double-slit structure), the interference pattern would be restored.  This erasure of quantum information restored the classical interference record.
Our proposal is similar to this, but is in some sense its opposite: we erase {\it classical} information in the measurement record, in order to restore the original {\it quantum} state.  Therefore, both the nature of the information that is being erased, and the goals of the information erasure are quite different.
 
We now give specific examples of weak continuous measurement in solid state physics to illustrate the physics of the ``quantum undemolition measurement''.

\begin{figure}[thb]
  \centering
 \includegraphics[width=14cm]{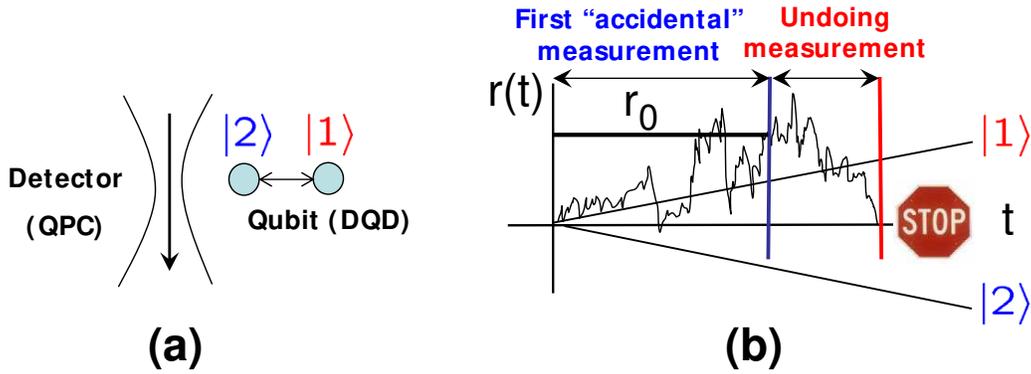} 
\caption{ (Modified after \cite{us}). {\bf (a)} Quantum point contact is capacitively coupled to a double quantum dot qubit.  The QPC current provides a continuous measurement of the quantum state of the DQD. {\bf (b)} Sample measurement record corresponding to a quantum undemolition measurement.  The initial measurement (registering a result $r_0$) at the blue line partially collapses the wavefunction.  The undoing procedure turns the detector back on, and waits until the measurement record crosses the origin (red line) where the detector is immediately turned off, uncollapsing the wavefunction. } 
\label{fig2}
\end{figure}

\section{Charge qubit}
\subsection{System characterization.}
A double-quantum-dot (DQD) qubit, measured continuously by a symmetric quantum point contact (QPC)  [Fig.~1(a)] has been extensively studied in earlier papers \cite{Gurvitz,weak-ss,DDexp}.  The qubit is formed from the lowest energy eigenstates in two nearby quantum dots. Quantum tunnelling between the dots lifts the degeneracy, and hybridizes the states, forming the two-state Hilbert space.  The quantum state may be described in the individual dot basis, where $|1\rangle$ corresponds to the right dot and $|2\rangle$ to the left dot ({\it c.f.} Fig.~1a).  The detector is the nearby quantum point contact, whose electron transmission properties are dependent on the DQD electron position via the Coulomb interaction. When electrical bias is applied across the QPC (that we take to be much larger than temperature or tunnel coupling energy), the measurement is characterized by the average currents $I_1$ and $I_2$ corresponding to the qubit states $|1\rangle$ and $|2\rangle$, and by the shot noise spectral density $S_I$ \cite{S-normalization}. We treat the additive detector shot noise as a Gaussian, white, stochastic process, and assume the detector is in the weakly responding regime, $|\Delta I| \ll I_0$, where $\Delta I=
I_1-I_2$ and $I_0=(I_1+I_2)/2$, 
so that the measurement process can be described by the quantum Bayesian formalism \cite{Kor-99}.
The continuous current from the QPC detector is then $I(t) = I_0 + (\Delta I/2) \langle \sigma_z \rangle + \xi$, where 
$\langle \xi(t) \xi(0)\rangle = (S_I/2) \delta(t)$.

\subsection{Measurement dynamics.} 
We assume for simplicity that there is no qubit Hamiltonian evolution
(this can also be effectively done using ``kicked'' quantum nondemolition (QND) measurements
\cite{Jordan}). As was shown in \cite{Kor-99}, the QPC is an ideal quantum
detector (which does not decohere the measured qubit), so that the evolution
of the qubit density matrix $\rho$ due to continuous measurement preserves
the ``murity'' ${\cal M}$ while the diagonal matrix elements evolve according to the classical Bayes rule.  We define the electrical current through the QPC in a time $t$ as $\bar{I}(t)=[\int_0^t I(t')\, dt']/t$, and together the quantum Bayesian equations read
\be
\rho_{11}(t) = \frac{\rho_{11}(0) P_1({\bar I})}{\rho_{11}(0) P_1({\bar I}) + \rho_{22}(0) P_2({\bar I})}, \qquad
\rho_{22}(t) = \frac{\rho_{22}(0) P_1({\bar I})}{\rho_{11}(0) P_1({\bar I}) + \rho_{22}(0) P_2({\bar I})}, \qquad
{\cal M} = \rho_{12}/\sqrt{\rho_{11}\rho_{22}} = {\rm const},
\ee
where the conditional (Gaussian) probability densities of a given current realization, given that the qubit is in $\vert 1 \ra, \vert 2\ra$ are 
\be
  P_{1,2}({\bar I})= \sqrt{t/\pi S_I}\, \exp [-({\bar{I}-I_{1,2}})^2 t/S_I].
\ee
These (nonlinear stochastic) equations may be simplified by noting 
     \begin{equation}
\frac{\rho_{11}(t)}{\rho_{22}(t)}=\frac{\rho_{11}(0)\exp [-(\bar{I}(t)-I_1)^2
t/S_I]}{\rho_{22}(0)\exp [-(\bar{I}(t)-I_2)^2 t/S_I]} =
\frac{\rho_{11}(0)}{\rho_{22}(0)}\, e^{2r(t)},
    \label{Bayes-DD}
    \end{equation}
where we define the {\it measurement result} as $r(t)=[\bar{I}(t)-I_0]\, t\Delta I/S_I$. For times
much longer than the ``measurement time'' $T_m= 2 S_I/(\Delta I)^2$ (the time
scale required to obtain a signal-to-noise ratio of 1), the average current
$\bar{I}$ tends to either $I_1$ or $I_2$ because the probability density
$P({\bar I})$ of a particular $\bar{I}$ is
    \be
    P({\bar I})=\sum\nolimits_{i=1,2}
 \rho_{ii}(0) P_i({\bar I}).
\label{output}
    \ee
Therefore $r(t)$ tends to $\pm \infty$, continuously collapsing the state to
either $|1\rangle$ (for $r\rightarrow \infty$) or $|2\rangle$ (for
$r\rightarrow -\infty$).   Importantly, for the special case when the initial state is pure, the state remains pure during the entire process.  This set of DQD measurement dynamics can be derived directly from the more general POVM formalism \cite{ourprb,Nielsen}. 

In order to describe how to uncollapse this wavefunction, we note that if $r(t)=0$ at some moment $t$, then the qubit state becomes exactly the same as it was initially, $\rho (t)=\rho(0)$.   This of course must be the case if $t=0$, {\it i.e.} before the measurement began, but is equally valid for some later time.  To see why this is so, we note that in the absence of noise, the measurement result from states $\vert 1\ra, |2\ra$ would simply be $r_{1,2}(t) =\pm t/T_m$.   With the noise present, the measurement outcome $r(t)=0$ splits the difference between states $\vert 1 \ra$ and $\vert 2\ra$.  Such an outcome corresponds to an equal statistical likelihood of the states $\vert 1\rangle$ and $\vert 2\rangle$, and therefore provides no information about the state of the qubit.  Another way of thinking about the undoing is the following: The detector will eventually give an unambiguous answer $r \rightarrow \infty$ or $r\rightarrow -\infty$.  One of the two measurements is giving a ``misleading'' answer: $r$ is going up when it should go down, or vice versa, while the other measurement is giving a ``true'' answer.  If we knew which was the misleading one (for example the first), we would have some information to make a valid inference about the quantum state.  However, there is no way of saying whether the first or the second is true, again leading us to the special ``no information'' state of knowledge.

\subsection{Uncollapsing the wavefunction for the charge qubit.}
Suppose the outcome of a measurement is $r_0$, partially collapsing the qubit state toward either state $\vert 1\ra$ (if $r_0 >0$), or state $\vert 2\ra$ (if $r_0 <0$).
The previous ``no information" observation suggests the following strategy: continue measuring, with the hope that after some time $t$ the stochastic result of the second measurement $r_u(t)$ becomes equal to $-r_0$, so the
total result $r(t)=r_0+r_u(t)$ is zero, and therefore the initial qubit state is fully restored. If this happens, the measuring device is immediately switched off and the undoing procedure is successful [Fig.~2(b)].
However $r(t)$ may never cross the origin, and then the undoing attempt fails.  

This strategy requires the observation of a particular (random) measurement result that may never materialize.  The strategy shifts the randomness to the amount of time that needs to elapse in order to find the desired current value.  Of course, in a given realization the current could take on the desired value multiple times, so we will take as our strategy to turn off the detector the first time the measurement result takes on $r=0$. This problem is known as a first passage process in stochastic physics, and there are many applications of this idea \cite{redner}.  A common example is a random walker near the edge of a cliff, and we are interested in the average time it takes before disaster strikes.

We may now apply the methods of first passage theory, to the wavefunction uncollapse problem \cite{us}.
The success probability $P_s$ for this procedure has been calculated to be 
\begin{equation}
P_s = e^{-|r_0|}/\left[ e^{r_0}\rho_{11}(0)+e^{-r_0}\rho_{22}(0)\right] ,
    \label{s}
    \end{equation}
and the mean waiting time $T_{\rm undo}$ until the measurement is undone is 
\be T_{\rm undo} = T_m  \, \vert r_0\vert. 
\label{wait} \ee 

The probability of success $P_s$ given by (\ref{s}) becomes very small for 
$|r_0|\gg 1$ (when the measurement result indicates a particular qubit state with good confidence), eventually
becoming $P_s=0$ for a projective measurement, recovering the traditional
statement of irreversibility in this limiting case.   Notice that while we began with a particular initial state, $\rho (0)$, the result (\ref{s}) applies to every quantum state, and therefore the QUD measurement succeeds or fails regardless of our knowledge of the initial state.

We again stress that the undoing of the measurement does not involve time-reversal: the QPC electrons that undo the measurement are {\it different} from the QPC electrons that ``do'' the measurement in the first place.

\begin{figure}[tb]
 \centering
\includegraphics[width=11cm]{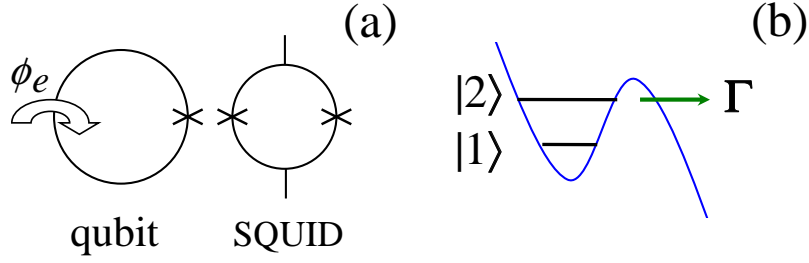}
\caption{(Modified after \cite{us}) {\bf (a)} Schematic of a phase qubit controlled by an external flux
$\phi_e$ and inductively coupled to the detector SQUID. {\bf (b)} Energy profile
$V(\phi )$ with quantized levels representing the qubit states. The
tunneling event is sensed by the SQUID.} 
\label{fig3}\end{figure}

\section{Phase Qubit}
\subsection{System Characterization.}
Our next example is for a superconducting ``phase'' qubit developed in the group of J. Martinis
\cite{Martinis,Katz}. The system (similar to the ``flux'' qubit) is comprised
of a superconducting loop interrupted by one Josephson junction [Fig.\
3(a)], which is controlled by an external flux $\phi_e$. Qubit states
$|1\rangle$ and $|2\rangle$ [Fig.\ 2(b)] correspond to the two lowest energy eigenvalues in
a quantum well with potential energy $V(\phi )$, where $\phi$ is the
superconducting phase difference across the junction. The qubit is measured by lowering the barrier (which is
controlled by $\phi_e$), so that the upper state $|2\rangle$ tunnels into
the continuum with rate $\Gamma$, while state $|1\rangle$ does not
tunnel out.  This state-selective measurement is possible because the tunneling rate is exponentially
sensitive to the width of the quantum barrier, so the lower energy level has a tunneling rate that is about a factor of 200 smaller than the upper level \cite{Katz}. The tunneling event is sensed by a two-junction detector SQUID
inductively coupled to the qubit [Fig.\ 3(a)]. 

\subsection{Measurement Dynamics}
    For sufficiently long tunneling time $t$, $\Gamma t \gg 1$, the measurement
corresponds to the usual collapse:  the qubit state is either projected onto the lower
state $|1\rangle$ (if no tunneling is recorded) or destroyed (if tunneling
happens). However, if the barrier is raised after a finite time $t \sim
\Gamma^{-1}$, the measurement is weak: the qubit state is still destroyed if tunneling happens, while in the case of no tunneling (a null-result measurement) the qubit density matrix evolves in the rotating frame as \cite{Katz}
    \begin{equation} 
    \rho_{11}(t) = \frac{\rho_{11}(0)}{\rho_{11}(0) + \rho_{22}(0)\, e^{-\Gamma t}}, \qquad
    \rho_{22}(t) = \frac{\rho_{22}(0)\,e^{-\Gamma t}}{\rho_{11}(0) + \rho_{22}(0)\, e^{-\Gamma t}}, \qquad
{\cal M}(t) = {\cal M}(0) e^{-i\varphi (t)},
    \label{Bayes-phase}
    \end{equation}
where the phase $\varphi (t)$ accumulates because of the adiabatic change of energy
difference between states $|1\rangle$ and $|2\rangle$ when the barrier
is lowered by changing $\phi_e$. Notice that except for the effect of the extra
phase $\varphi (t)$, the qubit evolution (\ref{Bayes-phase}) is similar to
the qubit evolution in the previous example; in particular, it also
represents an ideal measurement which does not decohere the qubit, and
has a clear Bayesian interpretation. Formally, the evolution
(\ref{Bayes-phase}) corresponds to the measurement result $r=\Gamma t/2$ in
Eq.\ (\ref{Bayes-DD}). The coherent non-unitary evolution
(\ref{Bayes-phase}) has been experimentally verified in Ref.\ \cite{Katz}
using tomography of the post-measurement state (in \cite{Katz} the product
$\Gamma t$ was actually varied by changing the tunneling rate $\Gamma$, while
keeping the duration $t$ constant).

\subsection{Uncollapsing the wavefunction for the phase qubit.}
A slight modification of the experiment \cite{Katz} can be used to
demonstrate measurement undoing.  Suppose the tunneling event did not happen
during the first weak measurement, so the evolution (\ref{Bayes-phase}) has
occurred. The undoing of this measurement consists of three steps:
$(i)$ Exchange the amplitudes of states $|1\rangle$ and
$|2\rangle$ by the application of a $\pi$-pulse, $(ii)$ perform another weak
measurement, identical to the first measurement, $(iii)$ apply
a second $\pi$-pulse. If the tunneling event did not occur during the second
measurement, then the information about the initial qubit state is erased (both
basis states have equal likelihood for two null-result measurements).
Correspondingly, according to Eq.\ (\ref{Bayes-phase}) (which is applied for
the second time with exchanged indices $1\leftrightarrow 2$), any initial
qubit state is fully restored (notice that the phase $\varphi$ is also
canceled).

\begin{figure}[tb]
 \centering
\includegraphics[width=14cm]{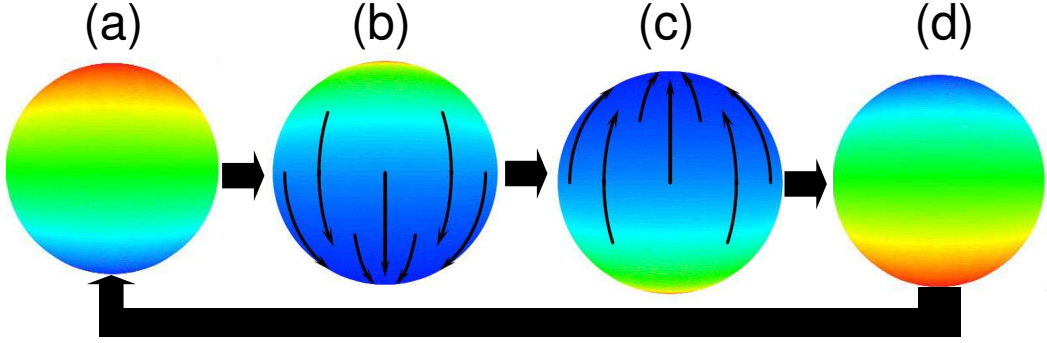}
\caption{Illustration of the QUD measurement for the phase qubit.  {\bf (a)} All possible qubit states are labeled by color on the Bloch sphere, with red corresponding to state $\vert 2\ra$ and blue to state $\vert 1\ra$. 
{\bf (b)}  The first weak measurement is accomplished by lowering the tunnel barrier for a time, $\Gamma t \sim 1$, shifting all states down toward the South pole (state $\vert 1\ra$) if no tunneling event is recorded. {\bf (c)} The next step is to apply a $\pi$-pulse to the qubit with an appropriately timed on-resonance microwave pulse, switching $\vert 1\ra \leftrightarrow \vert 2 \ra$. {\bf (d)} The undoing measurement is accomplished by lowering the tunnel barrier for the same time $\Gamma t$.  A final $\pi$-pulse restores the original orientation by again switching $\vert 1\ra \leftrightarrow \vert 2 \ra$.  If the second measurement also records no tunneling event, then the wavefunction is successfully uncollapsed.}
\label{fig4}\end{figure}

    The success probability $P_s$ for the undoing procedure is just the
probability that the tunneling does not happen during the second
measurement. If we start with the qubit state $\rho (0)$, the state after the
first measurement is given by Eq.\ (\ref{Bayes-phase}). After the
$\pi$-pulse, the occupation of the upper state is $\rho_{22}'=
\rho_{11}(0)/[\rho_{11}(0)+\rho_{22}(0)\,e^{-\Gamma t}]$, so the success
probability  $P_s=1-\rho_{22}' (1-e^{-\Gamma t})$ can be expressed as
    \begin{equation}
P_s= e^{-\Gamma t}/\left[ \rho_{11}(0)+e^{-\Gamma t}\rho_{22}(0)\right],
    \label{s-phase}
    \end{equation}
which formally coincides with Eq.\ (\ref{s}) for $r=\Gamma t/2$.
 
 An important issue is how one would experimentally check that the wavefunction is indeed uncollapsed. 
While measurement undoing is most important for an unknown state, in the demonstration experiment the initial state can be known, and tomography of the final state can be used to check that it is identical to the initial state.  Here tomography would involve three measurements:  the original measurement, the undoing measurement, and the tomography measurement, all separated by controlled unitary operations.

\section{Further Results}
We now briefly describe further results that extend and compliment the above analysis and discussion. 

\begin{itemize}
\item{By employing positive-operator valued measure (POVM) formalism, we have analyzed the general case of QUD measurement.  Given that the measurement is described by a measurement operator $M_r$ corresponding to the result $r$ acting on the initial unknown state, the inverse operator $M_r^{-1}$ cannot be applied deterministically, because a completeness relation must hold \cite{us}.  POVM formalism predicts the upper bound for the QUD success probability, and we have demonstrated that the two strategies outlined above saturate this upper bound.}
\item{Given $N$ coupled charge qubits, we have demonstrated an explicit general procedure of QUD measurement.  This is done with a QPC with a sufficiently strong nonlinearity that the QPC conducts only when all $N$ qubits are in state $|1\ra$. Using a sequence of $2^N$ operations, each involving {\it (i)} a $N$ qubit unitary rotation, {\it (ii)} measuring for a certain amount of time, and {\it (iii)} another $N$ qubit unitary, any measurement operation can be undone with maximal probability.}
\item{In the example of the quantum dot charge qubit, we have also examined the case where there is also Hamiltonian evolution of the quantum dot qubit during the continuous measurement.  In this case, the undoing requires additional single-qubit unitary operations before and after the continuous measurement, that corresponds to the singular-value decomposition of the inverse operation.}
\item{ Recently, a weak measurement implementation and QUD measurement was also proposed for a triplet-singlet quantum dot spin qubit \cite{spin}.  There, there is also state-selective tunneling of an electron on one of the dots, to the other dot.  The origin of this effect is {\it spin-blockade} - if the electrons are both spin up, the transition is energetically forbidden because the spin exchange energy puts the triplet level outside the transport window. If one electron is up and the other down, a singlet state can form, allowing the transition. }
\end{itemize}

\section{Conclusions}
Irreversibility has been considered one of the Hallmarks of quantum measurement.  We have shown that this property does not extend to {\it weak} measurements, and described experiments to demonstrate this phenomenon in solid-state quantum systems.  For the quantum dot charge qubit, a first-passage time strategy was employed, and for the superconducting phase qubit, an identical tunneling-time measurement was demonstrated. 
In quantum mechanics, reality and our knowledge about it seem to be intimately related.  The experiments proposed above clearly shows this connection in a more advanced and transparent setting than usual setting of ``strong'' collapse.  Therefore, even though our theory does not differ significantly in the philosophical sense from the standard Copenhagen interpretation, it shows the ``observer paradox'' in a new and significantly clearer way.  We conclude with a quotation from a recent {\it New Scientist} article on this subject, 
``Since the birth of quantum theory we have become used to thinking of quantum measurements as creating reality: until things are measured, they don't have an absolute, independent existence. But if some forms of measurement, such as weak measurement, are reversible, then the fundamentals of quantum mechanics go even deeper than we realized. If you create reality with weak quantum measurements, does undoing them erase the reality you created?'' \cite{ns}


\begin{thebibliography}{99}
\bibitem{boltz}  J. L. Lebowitz, Physics Today, {\bf 46}(9), 32 (1993).

\bibitem{ruelle}
D. Ruelle, Phys. Rev. Lett. {\bf 56}, 405 (1986).

\bibitem{peres}
A. Peres, Phys. Rev. A {\bf 30}, 1610 (1984); A. Peres, {\it Quantum
Theory: Concepts and Methods} (Kluwer, Dordrecht, 1995); 
P. R. Levstein {\it et al.}, J. Chem. Phys. {\bf 108}, 2718 (1998).


\bibitem{Neumann} J. von Neumann, {\it Mathematical Foundations of
        Quantum Mechanics\/} (Princeton Univ. Press, Princeton, 1955).

\bibitem{book}
B. H. Bransden and C. J. Joachain, {\it Introduction to Quantum Mechanics} (Prentice Hall, Longman, 1989)

\bibitem{wheeler}
J. A. Wheeler,  {\it Law Without Law}, in {\it Quantum Theory and Measurement}, Edited by J. A. Wheeler
and W. H. Zurek, (Princeton Series in Physics, Princeton University Press, 1983).

\bibitem{zurek}
see {\it e.g.} W. H. Zurek, Rev. Mod. Phys. {\bf 75}, 715 (2003) 

\bibitem{GRW}
G. C. Ghirardi, A. Rimini and T. Weber, Lett. Nuovo Cimento {\bf 27}, 293 (1980).

\bibitem{us}
A. N. Korotkov and A. N. Jordan, Phys. Rev. Lett. {\bf 97}, 166805 (2006).

\bibitem{weak-meas} E. B. Davies, {\it Quantum Theory of Open
    Systems} (Academic, London, 1976);
    M. B. Mensky, Phys. Usp. {\bf 41}, 923 (1998);
    C. M. Caves, Phys. Rev. D {\bf 33}, 1643 (1986);
    H. J. Carmichael, {\it An Open System Approach to Quantum Optics},
    Lecture notes in physics (Springer, Berlin,  1993);
    D. F. Walls and G. J. Milburn, {\it Quantum Optics}, (Springer, 2006).


\bibitem{Ueda-99} M. Koashi and M. Ueda, Phys. Rev. Lett. {\bf 82}, 2598
    (1999).

\bibitem{Ban-01} A. Ban, J. Phys. A {\bf 34}, 9669 (2001).

\bibitem{D'Ariano-03} G.M. D'Ariano, Fortschr. Phys. {\bf 51}, 318 (2003).

\bibitem{reversib} M. A. Nielsen and C. M. Caves, Phys. Rev. A {\bf 55}, 2547
    (1997); A. Royer, Phys. Rev. Lett. {\bf 73}, 913 (1994); {\bf 74},
    1040(E) (1995); M. Ueda, N. Imoto, and H. Nagaoka, Phys. Rev. A {\bf 53},
    3808 (1996); H. Mabuchi and P. Zoller, Phys. Rev. Lett. {\bf 76}, 3108
    (1996); F. Buscemi, M. Hayashi, M. Horodecki, arXiv:quant-ph/0702166.


\bibitem{Scully-82} M. O. Scully and K. Dr\"uhl, Phys. Rev. A {\bf 25}, 2208
(1982).

\bibitem{Gurvitz} S. A. Gurvitz, Phys. Rev. B {\bf 56}, 15215 (1997).

\bibitem{weak-ss} D. V. Averin, Fortschr. Physik {\bf 48}, 1055 (2000);
    H.-S. Goan and G. J. Milburn, Phys. Rev. B {\bf 64}, 235307 (2001);
    S. Pilgram and M. B\"uttiker, Phys. Rev. Lett. {\bf 89}, 200401 (2002);
    A. A. Clerk, S. M. Girvin, and A. D. Stone, Phys. Rev. B {\bf 67}, 165324
    (2003).

\bibitem{DDexp}
 T. Hayashi {\it et al.}, Phys. Rev. Lett. {\bf 91}, 226804 (2003);  J. M. Elzerman {\it et al.}, Phys. Rev. B {\bf 67}, 161308(R) (2003);
 J. R. Petta {\it et al.}, Phys. Rev. Lett. {\bf 93}, 186802 (2004).

\bibitem{S-normalization} We use normalization of the shot noise, in which
    $S_{I}=2eI(1-{\cal T})$, where ${\cal T}$ is the QPC transparency.

\bibitem{Kor-99} A. N. Korotkov, Phys. Rev. B {\bf 60}, 5737 (1999);
        Phys. Rev. B {\bf 63}, 115403 (2001).

\bibitem{Jordan} A. N. Jordan and M. B\"uttiker, Phys. Rev. B {\bf 71}, 125333 (2005);
A. N. Jordan, A. N. Korotkov, and M. B\"uttiker, Phys. Rev. Lett. {\bf 97}, 026805 (2006).

\bibitem{ourprb}
A. N. Jordan and A. N. Korotkov, Phys. Rev. B {\bf 74}, 085307 (2006).

\bibitem{Nielsen} M. A. Nielsen and I. L. Chuang, {\it Quantum computation
        and quantum information} (Cambridge University Press, Cambridge,
        2000).

\bibitem{redner}  S. Redner, {\it A Guide to First-Passage Processes} (Cambridge University Press, New York, 2001).

\bibitem{Martinis} J. M. Martinis, S. Nam, J. Aumentado, and C. Urbina,
Phys. Rev. Lett. {\bf 89}, 117901 (2002).

\bibitem{Katz} N. Katz {\it et al.}, Science {\bf 312}, 1498 (2006).

\bibitem{spin}
A. N. Jordan, B. Trauzettel, and G. Burkard,  arXiv:0706.0180  
    
\bibitem{ns}    
A. Gefter, New Scientist {\bf 194}(2603), 32 (2007).


\end{thebibliography}
\end{document}